\documentclass[structabstract]{aa}
\usepackage{natbib}
\include{biblio.tex}
\usepackage{txfonts}
\usepackage{graphicx}
\usepackage{supertabular}
\usepackage{longtable}
\usepackage{lscape}
\usepackage{hyperref}

\hypersetup{
    unicode=false,          
    colorlinks=true,       
    linkcolor=black,          
    citecolor=black,        
    filecolor=black,      
    urlcolor=blue           
} 

\begin{document}

\title{Analysis of hydrogen-rich magnetic white dwarfs detected in the Sloan Digital Sky Survey}
\titlerunning{Analysis of hydrogen-rich MWDs detected in the SDSS}

\author{B. K\"ulebi\inst{1}
  \and S. Jordan\inst{1}
   \and F. Euchner\inst{2}
 \and B. G\"ansicke\inst{3}
 \and H. Hirsch\inst{4}}

\offprints{B. K\"ulebi, \email{ bkulebi@ari.uni-heidelberg.de} }

\institute{Astronomisches Rechen-Institut, Zentrum f\"ur Astronomie der Universit\"at Heidelberg, M\"onchhofstrasse 12-14,
D-69120 Heidelberg, Germany \and Swiss Seismological Service, ETH Zurich, Sonneggstrasse 5, CH-8092 Zurich, Switzerland
\and Department of Physics, University of Warwick, CV4 7AL Coventry, United Kingdom \and Dr.-Remeis-Sternwarte, Sternwartstrasse 7, D-96049 Bamberg, Germany}


\abstract {A large number of newly discovered magnetic white dwarfs in the SDSS have so far only been analysed by visual comparison of the observations with relatively simple models for the radiation transport in a magnetised stellar atmosphere.}{We model the structure of the surface magnetic fields of the hydrogen-rich white dwarfs in the SDSS.} {We have calculated a grid of state-of-the-art theoretical optical spectra of hydrogen-rich magnetic white dwarfs with magnetic field strengths between 1 MG and 1200 MG for different angles between the magnetic field vector and the line of sight, and for effective temperatures between  7000\,K and 50000\,K. We used a least-squares minimization scheme with an  evolutionary algorithm in order to find the magnetic field geometry best fitting the observed data. We used simple centered dipoles or dipoles which were shifted along the dipole axis to model the coadded SDSS fiber spectrum of each object.} {We have analysed the spectra of all known magnetic DAs from  the SDSS (97 previously published plus 44 newly discovered) and also investigated the statistical properties of magnetic field geometries of this sample.}{The total number of known magnetic white dwarfs already more than tripled by the SDSS and more objects are expected from a more systematic search. The magnetic fields span a range between $\approx 1$ and 900 MG. Our results further support the claim that Ap/Bp population is insufficient in generating the numbers and field strength distributions of the observed MWDs, and either another source of progenitor types or binary evolution is needed. Moreover clear indications for non-centered dipoles exist in about $\sim$50\% of the objects which is consistent with the magnetic field distribution observed in Ap/Bp stars.}

\keywords{stars: white dwarfs -- stars: magnetic fields -- stars: atmospheres}
\maketitle 

\section{Introduction}
\label{Introduction}
White dwarfs with magnetic fields between $10^4$ and $ 10^9$ G are thought to represent more than 10\% of the total population of white dwarfs \citep{Liebertetal03}. The Sloan Digital Sky Survey (SDSS), the largest spectroscopic survey carried out to date, has discovered thousands of new white dwarfs, among them 102 with magnetic fields (MWDs) \citep{Gaensickeetal02, Schmidtetal03, Vanlandinghametal05}. By Data Release 3 (DR3) the number of known magnetic white dwarfs increased from 65 \citep{WickramasingheFerrario00,Jordan01} to 167 \citep{Kawkaetal07}. The first seven magnetic DAs (DAHs) uncovered from SDSS were identified visually in the area of the initial Early Data Release \citep[EDR][]{Gaensickeetal02}. \citet{Schmidtetal03} added 46 objects in the DR1, 38 of them DAH plus three new magnetic DB (DBH), and five new MWDs showing metallic and molecular lines. \citet{Vanlandinghametal05} reported on 49 additional new MWDs from the DR2 and DR3, specifically 46 new DAH, two new DQAs and one DQ with molecular bands.  

\citet{Schmidtetal03} and \citet{Vanlandinghametal05} determined the field strengths  and the  inclinations of magnetic dipoles by comparing visually the observed spectra  with model spectra. They have used an extension of the modeling method of \citet{LatterSchmidt87} and accounted for the effect of the change of magnetic field strength  on line depths and the variation of the field strength over the stellar surface for only the unpolarized radiation flux, namely Stokes parameter I. Their analyses with this simplified method of radiation transport resulted in dipolar field strengths for the  SDSS MWDs  between $1.5$\,MG and $\sim 1000$\,MG. Including the pre-SDSS, formerly known MWDs their sample consisted of 111 MWDs, 97 were classified as DAHs.

In this work we present the re-analysis of SDSS DAHs, published by \citet{Gaensickeetal02}, \citet{Schmidtetal03} and \citet{Vanlandinghametal05},  plus the analysis of 44 new ones from data up to DR7 (DR4 till DR7 were not systematically scanned for MWDs).

\section{SDSS data}
\label{SDSS data}
SDSS investigates five-band photometry of the Northern Galactic Polar Cap using the 2.5 meter telescope at Apache Point, New Mexico, with its special purpose instruments \citep{Fukugita96}. Follow-up spectroscopy of many stars is also performed with the twin dual beam spectrographs (3900\,-\,6200 and 5800\,-\,9200 \AA, $\lambda /\Delta \lambda \sim 1800$), in particular of blue objects like white dwarfs and hot subdwarfs \citep{Harrisetal03, Kleinmanetal04}. Since the energy distribution  of strongly magnetic white dwarfs can differ from nonmagnetic ones, MWDs are not only found in the SDSS color categories for white dwarfs or blue horizontal-branch stars, but may also fall into the color categories for quasars (QSOs), ``serendipitous blue objects'', and hot subdwarfs. Based on their colors, objects are assigned to fibers for follow-up spectroscopic investigations \citep[for spectroscopic target selection see ][]{Stoughton02}. 

In order to identify magnetic white dwarfs from these samples, different techniques were used. From the sample of white dwarfs, selected by color cuts in the $u$-$g$ vs $g$-$r$ color-color diagram, \citet{Gaensickeetal02} and \citet{Schmidtetal03} used visual inspection. In the work by \citet{Vanlandinghametal05} visual identification was augmented by the \textit{autofit} process \citep{Kleinmanetal04}, which fits spectra and photometry of hydrogen and helium white dwarfs to theoretical models. In particular white dwarfs with magnetic fields above 3\,MG, are flagged due to the poor $\chi^2$ fits of the \textit{autofit} process, therefore MWDs with weaker magnetic fields might be overlooked \citep{Vanlandinghametal05}.

In addition to the data from the former SDSS MWD papers (DR1-DR3), we have analyzed new data of nineteen additional objects from the HYPERMUCHFUSS \cite[HYPER velocity or Massive Unseen Companions of Hot Faint Under-luminous Stars Survey; see][]{Tillichetal09}. This survey aims at the detection of high velocity under-luminous B stars and white dwarfs. The candidates were chosen by the selection criterion ($u$-$g$)$<$0.4 and ($g$-$r$)$<$0.1 and spectral fits were performed in order to determine the radial velocity.
Some objects showed formally very high negative radial velocities  ($\le -100$\,km/s) but turned out to be DAHs. The reason for this is that the higher-order Balmer lines of magnetic white dwarfs are systematically shifted to the blue, even at relatively small magnetic fields ($\le 20$\,MG) due to the quadratic Zeeman effect, mimicking a high radial velocity.  Additionally, 34 DAHs were serendipitously found in the course of a visual inspection of blue stellar objects from DR7. The total number of DAHs from SDSS is likely to grow further once a systematic search through all DR7 spectra is carried out. 

The one-dimensional spectra which we used in this work  were generated by SDSS's spectroscopic pipeline \texttt{spectro2d} and downloaded from the Data Archive Server.

\begin{table*}
\caption{Photometric properties of the new confirmed DAHs and their temperatures. The columns indicate the SDSS name of the object; the plate, Modified Julian Date and fiber ids of the observations; the SDSS photometric magnitudes $u$, $g$, $r$, $i$, $z$; and finally the temperatures derived from their colors.}\label{photometry}
\begin{tabular}{lccccccl}
\hline
\hline
MWD (SDSS+) & Plate-MJD-FiberID &$u$ / mag  &$g$ / mag &$r$ / mag &$i$ / mag & $z$ / mag & $T_{\rm eff}$ / K\\
\hline
J023420.63+264801.7 & 2399-53764-559 & 18.70 & 18.38 & 18.59 & 18.76 & 19.03 & 13500 \\
J031824.19+422651.0 & 2417-53766-568 & 18.59 & 18.23 & 18.32 &  18.43  & 18.67 & 10500  \\
J032628.17+052136.3 & 2339-53729-515 & 18.69 & 18.93 & 19.30 & 19.60 & 19.61 & 25000 \\
J033320.36+000720.6 & 0415-51879-485 & 17.07 & 16.52 & 16.39 & 16.35 & 16.44 & 7000\footnotemark[1]\footnotemark[2] \\
J074924.91+171355.4 & 2729-54419-282 & 18.78 & 18.78 & 19.13 & 19.44 & 19.64 & 20000 \\
J075234.96+172525.0 & 1920-53314-106 & 18.78 & 18.44 & 18.44 & 18.50 & 18.64 & 9000  \\
J080359.93+122943.9  &  2265-53674-033 & 17.24 & 17.23 & 17.53 & 17.83 & 18.08 & 9000 \\
J081716.39+200834.8  &  2082-53358-444 & 18.91 & 18.34 & 18.15 & 18.12 & 18.23 & 7000 \\
J083448.63+821059.1  & 2549-54523-135 & 18.07 & 18.32 & 18.74 & 19.06 & 19.49 & 27000 \\
J083945.56+200015.7 & 2277-53705-484 & 18.11 & 17.83 & 18.11 & 18.36 & 18.66 & 15000  \\
J085106.12+120157.8 & 2430-53815-229 & 17.35 & 16.96 & 17.14 & 17.30 & 17.56 & 11000  \\
J085523.87+164059.0 & 2431-53818-522 & 18.78 & 18.55 & 18.80 & 19.05 & 19.32 & 15500 \\
J085550.67+824905.3  & 2549-54523-066 & 18.40 & 18.60 & 18.91 & 19.23 & 19.46 & 25000 \\
J091005.44+081512.2  &  1300-52973-639 & 17.38 & 17.54 & 17.96 & 18.28 & 18.65 & 25000 \\
J091833.32+205536.9 & 2288-53699-547 & 18.73  & 18.41  & 18.66 & 18.92 & 19.22 & 14000  \\
J093409.90+392759.3  &  1215-52725-241 & 18.72 & 18.35 & 18.40 & 18.50 & 18.55 & 10000 \\
J094235.02+205208.3 & 2292-53713-019 & 18.41 & 18.42 & 18.80 & 19.05 & 19.26 & 20000 \\
J100657.51+303338.1 & 1953-53358-415 & 19.22 & 18.83 & 18.90 & 19.04 & 19.18 & 10000  \\
J100759.80+162349.6  & 2585-54097-030 & 18.01 & 17.70 & 17.80 & 17.96 & 18.19 & 11000 \\
J101428.09+365724.3 & 52993-1426-021 & 19.26 & 18.87 & 18.97 & 19.09 & 19.43 & 10500 \\
J102220.69+272539.8  &  2350-53765-543 & 20.47 & 20.05 & 20.16 & 20.38 & 20.69 & 11000 \\
J102239.06+194904.3 & 2374-53765-544 & 19.43 & 19.01 & 19.01 & 19.11 & 19.13 & 9000  \\
J103532.53+212603.5  &  2376-53770-534 & 17.98 & 17.40 & 17.23 & 17.19 & 17.21 & 7000\footnotemark[2] \\
J105709.81+041130.3  &  0580-52368-274  & 18.09 & 17.67 & 17.58 & 17.60 & 17.70 & 8000 \\
J112030.34-115051.1  & 2874-54561-512  &  18.65	& 18.73	& 19.05	& 19.34	& 19.75 & 20000\\
J112257.10+322327.8 & 1979-53431-512 & 19.60 & 19.37 & 19.50 & 19.68 & 19.92 & 12500  \\
J112328.49+095619.3  &  1222-52763-625 & 18.15 & 17.70 & 17.74 & 17.87 & 18.02 & 9500 \\
J113215.38+280934.3  &  2219-53816-329 & 17.50 & 16.99 & 16.88 & 16.87 & 16.92 & 7000\footnotemark[2] \\
J124836.31+294231.2 & 2457-54180-112 & 18.44 & 17.80 & 17.59 & 17.54 & 17.56 & 7000\footnotemark[2] \\
J125434.65+371000.1 & 1989-53772-41 & 16.01 & 15.97 & 16.35 & 16.64 & 16.95 & 10000  \\
J125715.54+341439.3 & 2006-53476-332 & 17.14 & 16.78 & 16.81 & 16.92 & 17.11 & 8500  \\
J134820.79+381017.2 & 2014-53460-236 & 17.26 & 17.54 & 18.04 & 18.33 & 18.70 & 35000  \\
J140716.66+495613.7 & 1671-53446-453 & 19.03 & 19.13 & 19.43 & 19.75 & 19.97 & 20000  \\
J141906.19+254356.5 & 2131-53819-317 & 17.80 & 17.41 & 17.46 & 17.53 & 17.69 & 9000 \\
J143019.05+281100.8 & 2134-53876-423 & 18.03 & 17.68 & 17.68 & 17.74 & 17.92 & 9000 \\
J151130.17+422023.0 & 1291-52738-615 & 18.20  & 17.98  & 18.01 & 18.20 & 18.48 & 9500  \\
J151415.65+074446.5  &  1817-53851-534 & 19.16 & 18.84 & 18.88 & 18.99 & 18.88 & 10000 \\
J152401.60+185659.2 & 2794-54537-410 & 18.39 & 18.15 & 18.34 & 18.54 & 18.8 & 13500 \\
J153843.10+084238.2  & 1725-54266-297 & 18.24 & 17.90 & 17.94 & 18.22 & 18.20 & 9500 \\
J154305.67+343223.6 & 1402-52872-145 & 18.08 & 18.32 & 18.75 & 19.10 & 19.46 & 25000 \\
J165249.09+333444.9 & 1175-52791-095 & 19.11 & 18.63 & 18.63 & 18.65 & 18.92 & 9000 \\
J202501.10+131025.6 & 2257-53612-167 & 18.91  & 18.76  & 19.07 & 19.28 & 19.74 &17000 \\
J220435.05+001242.9 & 0372-52173-626 & 19.66 & 19.38 & 19.47 & 19.54 & 19.71 & 22000 \\
J225726.05+075541.7 & 2310-53710-420 & 17.09  & 17.11  & 17.31 & 17.44 & 17.65 & 40000 \\
\hline
\end{tabular}
\label{tab:newMWD}
\\
${}^1$ HE 0330-0002\\
${}^2$ The temperature from fits to the color-color diagram is uncertain.
\end{table*}

\section{Analysis}
\label{Analysis}

\begin{figure*}
   \centering
   \resizebox{!}{0.93\textheight}{\includegraphics[viewport=45 39 595 812]{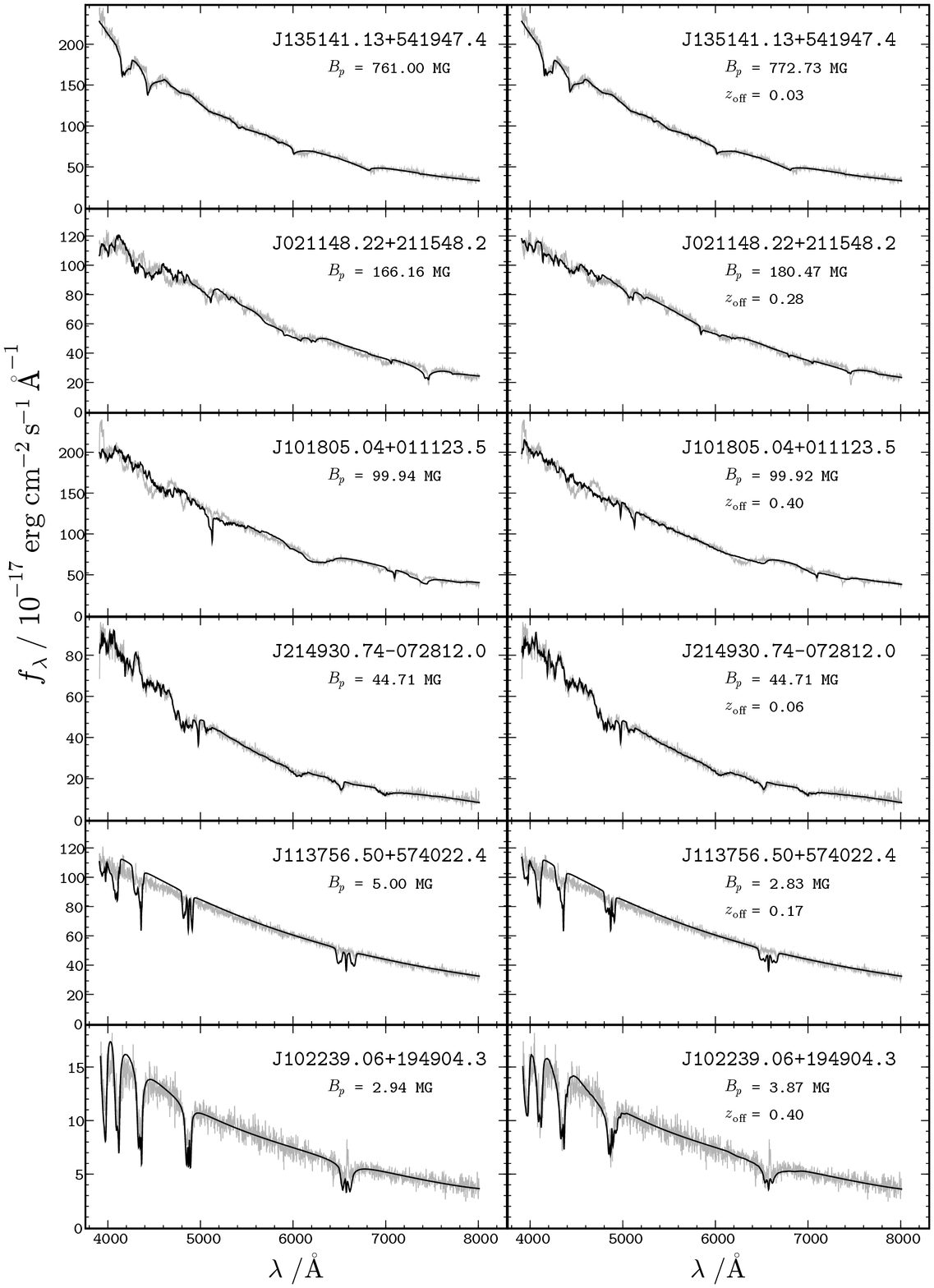}}
  \caption{Fits of observed spectra of DAHs from the SDSS to centered magnetic dipoles with 
   a polar field strength  $B_p$ (left) and dipoles shifted by $z_{\rm off}$ stellar radii along the dipole axis
  (right). Representative fits and objects mentioned throughout the article are chosen. The color version of this figure and the remaining 128 fits can be found in the online version of this paper, \href{http://www.ari.uni-heidelberg.de/mitarbeiter/bkulebi/papers/12570_online.pdf}{here}.} 
  \label{fig:bestfit}
\end{figure*}

Our model spectra are calculated with a radiative transfer code for magnetized white dwarf atmospheres, which for a given temperature and pressure structure of a model atmosphere ($T_{\rm eff}$, $\log g$) and  a given magnetic field vector with respect to the line of sight and the normal on the surface of the star, calculates theoretical flux and polarization spectra \cite[see][]{Jordan92,JordanSchmidt03}.

In order to increase efficiency, we pre-computed a three-dimensional grid of Stokes $I$ and $V$ ($V$ spectrum not used due to the lack of polarization measurements) model spectra with  effective  temperature $7000\,{\rm K}\le T_{\rm eff} \le 50000\,{\rm K}$ in 14 steps, magnetic field strength $1\,{\rm MG}\le B\le 1.2\,{\rm GG}$ in 1200 steps, and 17  different  directions of $\psi$ relative to the line of sight as the independent variables (9 entries, equally spaced in   $\cos \psi$). All spectra were calculated for a surface gravity of \mbox{$\log g = 8$}. Since no polarization information is available from the SDSS, our analysis is limited to the flux spectra (Stokes parameter $I$). Limb darkening is accounted for by a simple linear scaling law\citep[see][]{Euchneretal02}.

The magnetic field geometry of the DAHs was determined with a modified version of the code developed by \citet{Euchneretal02}. This code calculates the total flux (and circular polarization) spectra for an arbitrary magnetic field topology by adding up appropriately weighted model spectra for a large number of surface elements and then evaluating  the goodness of fit. Magnetic field geometries are accounted for by  multipole expansions of the scalar magnetic potential. The individual multipole components may be independently oriented with respect to the rotation axis of the white dwarf and offset with respect to its center, allowing in principle for rather complex surface field topologies. Additional free parameters are the white dwarf effective temperature and the inclination of the rotation axis with respect to the line of sight. Observed spectra can be fitted using an evolutionary algorithm  \citep{Rechenberg94} with a least-squares quality function.

Additionally to the Zeeman effect, Stark broadening has to be considered. For the case when the electric and magnetic fields are parallel \citet{Friedrichetal94} have estimated the effect on stationary line components, which are  transitions that vary slowly in wavelength for large intervals of magnetic field strengths. Stationary lines are more pronounced than non-stationary lines, since they are not smeared out extensively due to the variation of the magnetic field strength over the stellar surface.

However, no atomic data for hydrogen in the presence of both a magnetic and electric field are available for arbitrary strengths and arbitrary angles between two fields. Therefore, only a crude approximation  \cite[see ][]{Jordan92} is used in our model and systematic uncertainties are unavoidable, particularly in the low-field regime ($\le 5$\,MG) where the Stark effect dominates.  Consequently, effective temperatures and surface gravities derived from fitting the Balmer lines alone are less reliable than in the case of non-magnetic white dwarfs. This may also result in disagreements with temperature estimates derived from the continuum slope.

\begin{figure*}
   \centering
   \resizebox{!}{0.93\textheight}{\includegraphics[viewport=45 39 595 812]{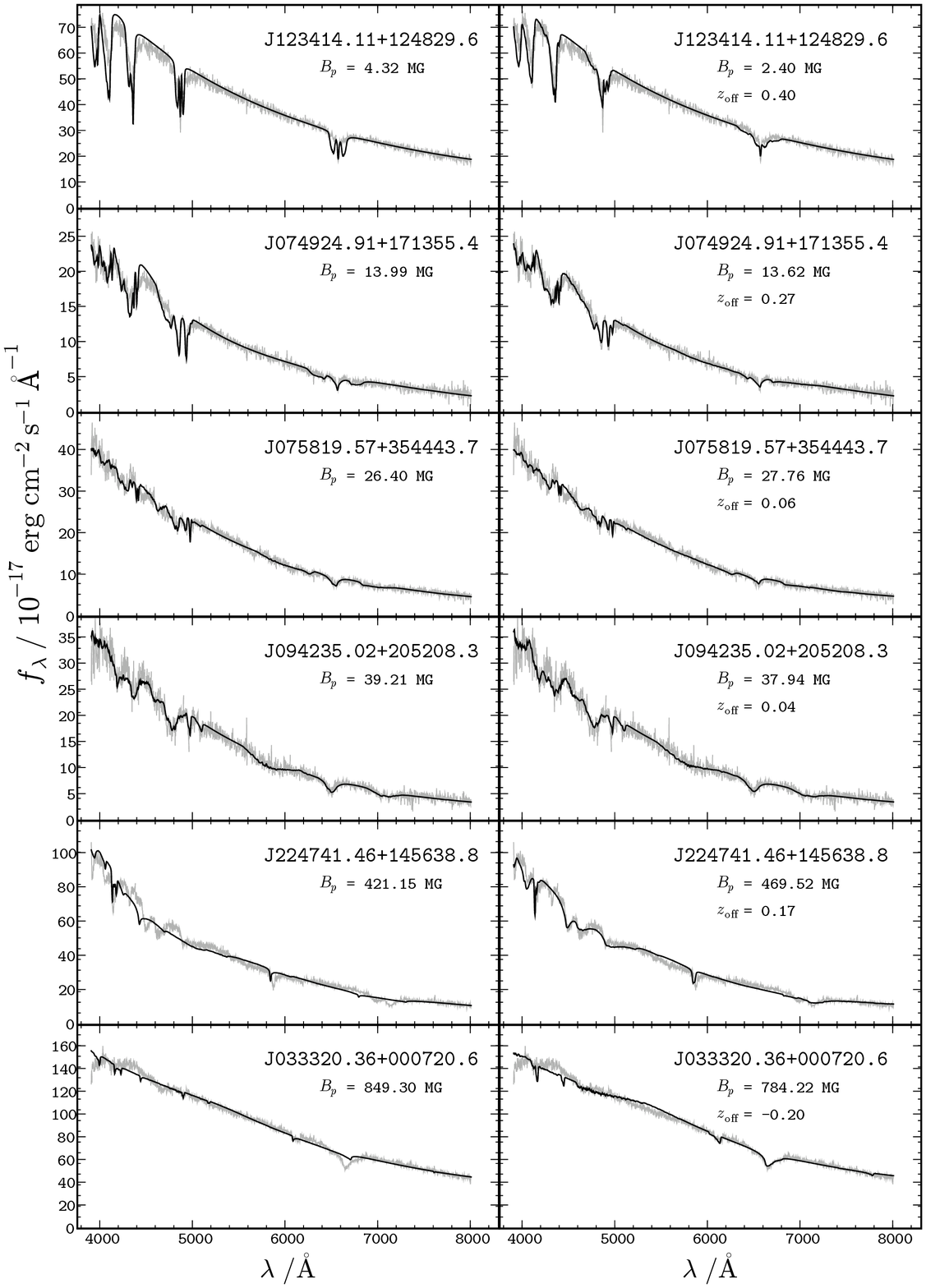}}
  \caption{Fits of observed spectra of DAHs from the SDSS to centered magnetic dipoles with 
   a polar field strength  $B_p$ (left) and dipoles shifted by $z_{\rm off}$ stellar radii along the dipole axis
  (right). Representative fits and objects mentioned throughout the article are chosen. The color version of this figure and the remaining 128 fits can be found in the online version of this paper, \href{http://www.ari.uni-heidelberg.de/mitarbeiter/bkulebi/papers/12570_online.pdf}{here}.}
  \label{fig:bestfit02}
\end{figure*}

Time-resolved analysis for rotating single magnetic white dwarfs was instrumental in determining rather complex field structures \cite[e.g. VLT observations by][]{Euchneretal02,Euchneretal05,Euchneretal06}. However this usually relies on the preliminary knowledge of period which is usually derived via photometry, separately. Although the individual SDSS fiber spectra exists with 15 minute exposure time, due to the lack of information on spin period, we constrained ourselves to the coadded spectra which includes 3 or more individual spectra with total exposure time of at least 45 minutes. With the possible exception of a few bright objects, the signal to noise of the individual spectra would not be sufficient to find indication for rotational changes. Therefore, we had to restrict ourself to simple models for the magnetic field geometry, namely centered magnetic dipoles with only two free parameters or to dipoles offset along the magnetic axis which have three free parameters. These parameters are the magnetic dipole field strength $B_p$,and the inclination of the dipole axis $i$ for centered dipole. For the offset dipole there is an additional offset parameter along the magnetic axis $z_{\rm off}$ in terms of the stellar radius. For the 97 DAHs  analyzed, we used the literature values for $T_{\rm eff}$ which were determined by comparison to the theoretical non-magnetic DA colors in the $u$-$g$ vs $g$-$r$ plane \citep{Schmidtetal03,Vanlandinghametal05}. The temperature of the new DAHs presented in Table\,\ref{tab:newMWD} were estimated by the synthetic SDSS color-color diagrams by \citet{HolbergBergeron06}\footnote{http://www.astro.umontreal.ca/$\sim$bergeron/CoolingModels/}, assuming that the influence of the magnetic field at the temperature determination is small, which is not always the case \cite[see][and Sec.\ref{Geometry}]{Schmidtetal86,Gaensickeetal01}.

All fits have reduced  $\chi^2$ values between 0.8 and  3.0 except for some high-field objects which obviously deviate from the assumed dipole geometry (see Sect.\,\ref{Results}). We use the error calculation method of \citet{Zhang86}, which assumes that a small change in $\chi^2$ could be approximated by a linear expansion of the covariance matrix; for  complex $\chi^2$ topologies, this approximation is not sufficient. Moreover, the final error for the inclination is often very large. 

Final fit parameters with errors are noted in Table\,\ref{tab:results}. In Fig.\,\ref{fig:bestfit} and \ref {fig:bestfit02} we have shown fits of 12 DAHs, as an example. 
All of our remaining resulting fitted spectra can be found in the online version of this  article (Fig.\,A.3-A.23 for the  other DAHs).

\begin{figure}
   \resizebox{0.97\hsize}{!}{\includegraphics{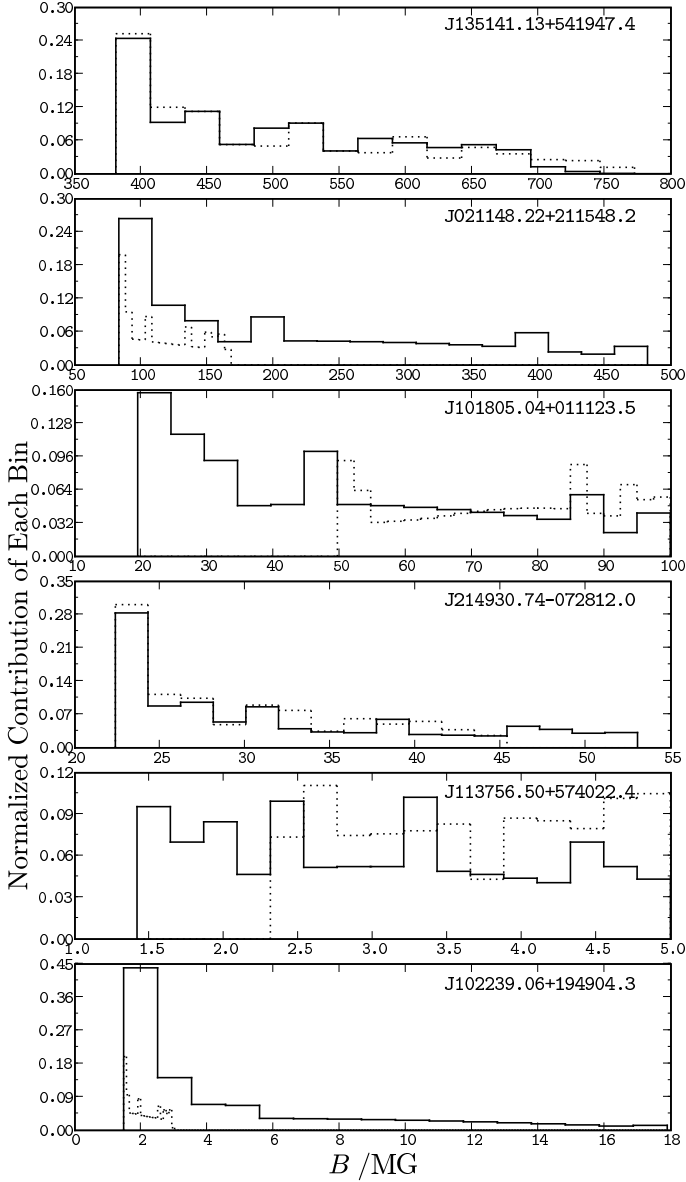}}
  \caption{Normalized histograms of the magnetic field strength distributions over the visible hemisphere of the star used for calculation of the synthetic spectra shown on Fig.\,\ref{fig:bestfit}. Dotted lines represent the centered dipole models, solid lines indicate dipole models with offsets.}
  \label{fig:b_histograms}
\end{figure}


\section{Results}
\label{Results}

\subsection{Individual objects}
\label{Individual}
Three objects analyzed by \citet{Schmidtetal03} and \citet{Vanlandinghametal05} are omitted in this work. \textit{SDSSJ05959.56+433521.3} (\textit{G111-49}) was listed by \citet{Schmidtetal03} as a DAH, but is a Carbon-rich (DC) MWD \citep{Putney95}. \textit{SDSSJ084716.21+148420.4} is a DAH+DB binary, in which the Helium component is quite strong in the spectrum. This dilution of the Hydrogen features hindered the analysis of this object by our code. Finally we failed to model \textit{J220029.08-074121.5} due to the lack of any discernible features in the SDSS spectrum.

Emission lines were found in \textit{SDSSJ102220.69+272539.8} and \textit{SDSSJ\-102239.06\-+194904.3}  (the latter is shown in  Fig.\,\ref{fig:bestfit}), very similar to \textit{SDSSJ121209.31\-+013627.7} which could indicate that these objects may be EF Eri like, magnetic cataclysmic variables with a brown dwarf companion \citep{Schmidtetal05, Debesetal06, Burleighetal06, Farihietal08}.

The spectra of the high-field objects \textit{SDSSJ224741.41\-+145638.8} and \textit{SDSSJ101805.04\-+01123.5} (PG\,1015+014, shown in  Fig.\,\ref{fig:bestfit}) do not fit particularly well. At higher field strengths ($>\,50$\,MG) the spectra become very sensitive to the details of the magnetic field geometry, as was demonstrated by \citet{Euchneretal02,Euchneretal05,Euchneretal06}.  The deviations of the observed spectra from our theoretical spectra assuming  (offset) dipole models hint therefore to a magnetic field geometry that is more complex than a shifted dipole. A more comprehensive analysis of \textit{SDSSJ101805.04\-+01123.5} showed that individually tilted and off-centered zonal multipole components with field strengths between 50-90 MG is needed to represent the global magnetic field \citep{Euchneretal06}, which was consistent with our analysis. 

The colors of MWDs with high field strengths ($>\,50$\,MG) are known to differ from non-magnetic white dwarf colors due to the absorption from their spectral features as noted in Sec.\,\ref{SDSS data}. This behavior also affects temperature determinations from color-color diagrams. The analysis of \textit{SDSSJ224741.41\-+145638.8} by \citet{Euchneretal06} revealed effective temperature of $T_{\rm eff}=10000\pm1000$\,K unlike $T_{\rm eff}=12000$\,K  that is derived from color-color diagrams. We have used 10000\,K for our models and this value gave better results, especially on the basis of line depths. When color derived effective temperatures were used a similar discrepancy with the slopes and line depths was also observed in \textit{SDSSJ224741.41\-+145638.8}. 17000\,K was used in modeling by \citet{Schmidtetal03}, but the colors of \textit{SDSSJ224741.41\-+145638.8} lies beyond the \citet{HolbergBergeron06} grid of log$g$ - $T_{\rm eff}$ in $u$-$g$ -- $g$-$r$ plane. In our procedure we accomplished the best result with 50000\,K for this object on the basis of slope and line depths. On the other hand, some high-field objects in our sample like \textit{SDSSJ135141.13\-+541947} (Fig.\,\ref{fig:bestfit}) were fitted well and this discrepancy between the temperature derived by colors vs spectral fits was not observed.

For some fits, observed and computed spectra strongly differ in line depths. These unsatisfactory fits revealed two different kind of symptoms: Either $\sigma^{\pm}$ components were shallower than expected from their sharp $\pi$ counterparts in observed spectra with respect to the models; or both $\sigma^{\pm}$ and $\pi$ components of the lines were shallow at the same time.

\begin{figure}
   \resizebox{0.97\hsize}{!}{\includegraphics{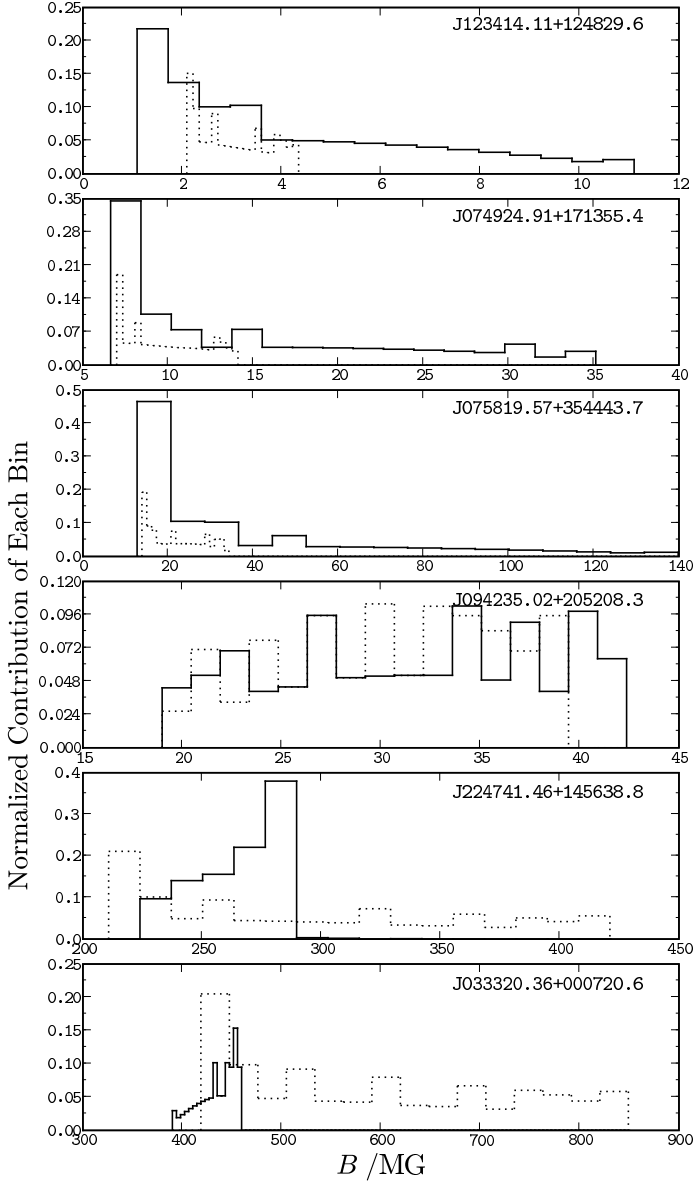}}
  \caption{Normalized histograms of the magnetic field strength distributions over the visible hemisphere of the star used for calculation of the synthetic spectra shown on Fig.\,\ref{fig:bestfit02}. Dotted lines represent the centered dipole models, solid lines indicate dipole models with offsets.}
  \label{fig:b_histograms02}
\end{figure}

\begin{figure}
   \centering
   \resizebox{\hsize}{!}{\includegraphics{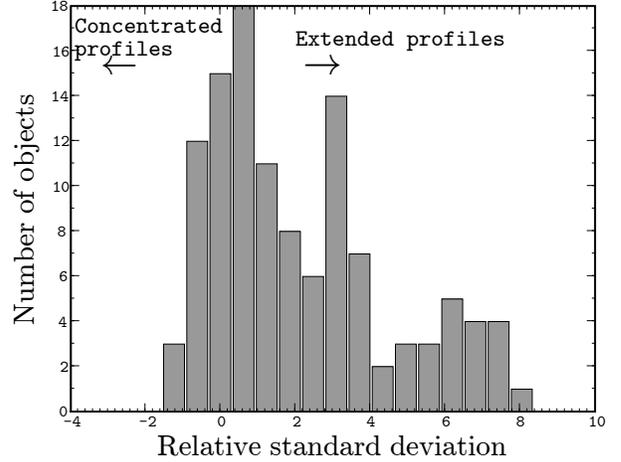}}
  \caption{Histogram of the relative variance $\sigma_{\rm rel}$ (see Sec.\,\ref{Geometry}) for known SDSS MWDs. Negative $\sigma_{\rm rel}$ indicates a magnetic field strength profile of the visible surface which is more concentrated than a centered dipole, whereas positive indicates profiles that are more extended.}
  \label{fig:relative_std}
\end{figure}

The occurrence of sharp line cores with shallow wings in the spectra was already noted for \textit{J123414.11+124829.6} by \citet{Vanlandinghametal05}. It has been suggested that this might be a result of deviation from centered dipole geometry, and our fits with offset dipole models proved to be considerably better than the centered dipole models in reduced $\chi^2$. For lower magnetic strengths ($<\,50$\,MG) the smearing effect of offset dipole models affects only the $\sigma^{\pm}$ components of the lines. The reason is that in this field regime $\sigma^{\pm}$ components become more separated while the $\pi$ components are only slightly blue shifted with increasing field strength. Therefore smeared out wings with sharp line cores can be synthesized by adding up a larger range of magnetic field strength values. Dipole models with offsets can generate such extended magnetic field distributions (see Sec.\,\ref{Geometry}). Our fit to \textit{J123414.11+124829.6} was considerably better but we did not reproduce the exact profile. Another possible explanation of such a spectrum is the contribution from a non-magnetic DA, which would dilute the $\sigma^{\pm}$ components causing an increased contrast between wings and line cores. dipole models improved our fits, further analysis is needed to disentangle the effect of geometry versus possible contribution from a non-magnetic DA. 

For the other case, \textit{J113756.50+574022.4} (see Fig.\,\ref{fig:bestfit}) has very shallow features, with discernible magnetic wings. Neither a complicated geometry nor a change in effective temperature explains this lack of line depth. Nevertheless, the magnetic field strength could be derived from the extent of the wings. We suspect that this object may be an unresolved spectroscopic binary (e.g. with  DA+DC components), since in these situations Hydrogen line strengths are known to be suppressed by the other component \citep{Bergeron90, LiebertBergeron93}. The other objects with shallow features which belong to this category are: \textit{J084716.21+484220.4}, \textit{J090632.66+\-080716.0}, \textit{J113215.38+\-280934.3}, \textit{J103532.53+\-212603.5},  \textit{J112328.49+\ 095619.3}, \textit{J124806.38+\-410427.2}, \textit{J141906.19+\-254356.5}. 

One interesting outcome of our work was the modeling of \textit{J033320.36+000720.6} which was formerly identified in the Hamburg/ESO survey for bright quasars, as HE 03330-0002 \citep{Reimersetal98}. Its magnetic property was confirmed by \citet{Schmidtetal01} by circular polarimetry; nevertheless the modeling of the HE 03330-0002 was not possible, since the transitions in its spectrum were thought not to be explainable with hydrogen or with helium transitions. Although \textit{J033320.36+000720.6} was already discovered in the EDR, due to this lack of knowledge about the atmosphere \citet{Gaensickeetal02} did not try to model it with pure hydrogen atmosphere. 

However we have noticed that some of the lines in the spectrum of \textit{J033320.36+000720.6} possibly coincided with hydrogen stationary lines. With this incentive we modeled it with a 7000\,K DAH atmosphere. Our initial fits with a centered dipole resulted in a dipolar field strength of $\sim$850 MG (see Fig\,\ref{fig:bestfit02}) which produced the position but not the depths of three transitions. More careful modeling with an offset dipole showed that the mean field strength over the visible surface is much more concentrated than a regular dipole field geometry. Fig.\,\ref{fig:b_histograms02} shows that the mean field strength is dominated by an interval of field strengths between $390-470$ MG. When we consider the apparent peak value of this distribution, we get commensurable line positions for three lines in the spectrum as H$\alpha$ transitions (see table\,\ref{tab:03330_lines}).

The pure hydrogen atmosphere did not account for the red part ($\lambda < 5500 \AA$) of the spectrum. Apparently there are additional opacity contributions from different species of elements, both to continuum and to the line features. 

\begin{table}
\caption{The hydrogen transitions of SDSSJ033320.36+000720.6 and their wavelengths at $446.5$ MG.}\label{tab:03330_lines}
\begin{tabular}{llr}
\hline
\hline
Line & $nlm-n^{\prime} l^{\prime }m^{\prime}$ &$\lambda (\AA)$\\\hline
H$\alpha$& $2s_0\ \ -\,3p_0$ & 6113.05 \\
H$\alpha$ & $2s_0\ \ -3p_{-1}$ & 7710.91  \\
H$\alpha$ & $2p_{-1}-3d_{-2}$ & 6647.76 \\
\hline
\end{tabular}
\label{tab:03330_lines}
\\
\end{table}

\subsection{Magnetic field geometry}
\label{Geometry}
In this work we have modeled a large sample of DAHs with dipole magnetic fields with offsets of the magnetic axis. However, visualization of these model parameters is not straightforward. The effect of the dipole offset on the model spectrum depends on both inclination and the polar field strength. However, the polar field strength is not a representative value for the global magnetic field on the visible surface if the offsets are large. The most direct way to investigate a model geometry is to construct a diagram with the angle between the line of sight and the local magnetic field vector versus the magnetic field strength plots which are equivalent to the ZEeman BRoadening Analysis (ZEBRA) plots of \citep{Donati94}. For our case since we did not have polarization data, we only considered the magnetic field strength distribution histograms for simplicity.

In general the effect of the offset dipole models is to extend or reduce the range of magnetic field strengths over the visible surface of the MWD, which is a fixed factor of two for centered dipole models (see Fig.\,\ref{fig:b_histograms}). For offset dipoles the range depends on the values of $B_p$ and the inclination $i$. In order to quantify the difference between the centered and off-centered dipole models
we determined the average and  the standard deviation $\sigma$  of the distribution of the magnetic fields for the parameters of our best fits. The relative change of the standard deviations $\sigma_{\rm centered}$ and $\sigma_{\rm offset}$ is given by $\sigma_{\rm rel} = \frac{\sigma_{\rm offset}- \sigma_{\rm centered}}{\sigma_{\rm centered}}$. 

$\sigma_{\rm rel}=0$ indicates that the width of the centered and offset dipole  models are the same;  $\sigma_{\rm rel}<0$ means more concentrated than a dipole field, in the case of $\sigma_{\rm rel}>0$ the distribution of field strengths is more extended. 

To discuss the magnetic field geometry of our sample, we plotted the histogram of  $\sigma_{\rm rel}$ values for all known SDSS DAHs except the ones that were discussed in Sec.\,\ref{Individual} as possible binaries (Fig.\,\ref{fig:relative_std}). The average $\sigma_{\rm rel}$ for this sample turned out to be $2.18$.
This means even with rather mediocre signal to noise spectra of SDSS, an overall the tendency towards non-dipolarity is observed in our sample of white dwarfs.

\begin{figure}
   \centering
   \resizebox{\hsize}{!}{\includegraphics{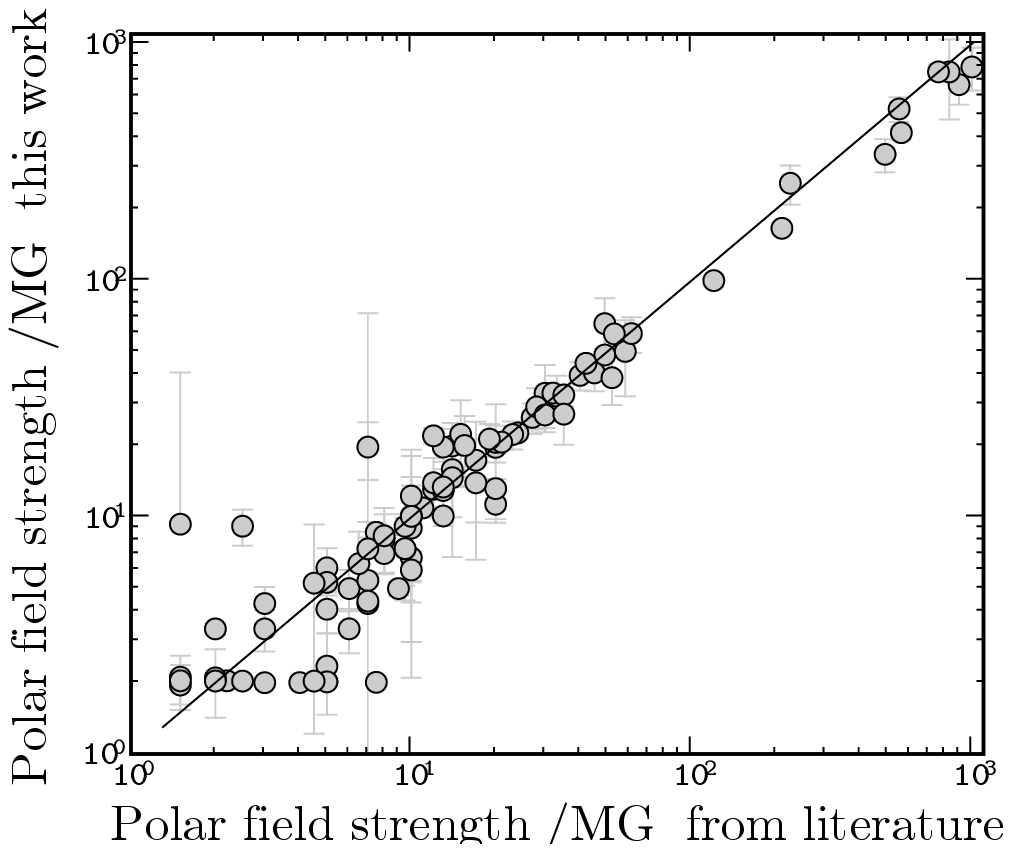}}
  \caption{Comparison of centered dipole magnetic field fit values in this work versus \citet{Schmidtetal03}, \citet{Vanlandinghametal05}.}
  \label{fig:comparison}
\end{figure}

\subsection{General discussion}
\label{General discussion}
Overall our results are consistent with the former analyses on DAHs (see Fig.\,\ref{fig:comparison}), which shows that simple atmosphere models with pre-assumed dipole magnetic values are good approximations for these objects. In all of the cases offset dipole models resulted in significantly better fits than the models with centered dipoles. We have noted in Section \ref{Individual} that still for some DAHs with high fields, completely satisfactory fits cannot be achieved with offset dipole models. This hints to a magnetic field geometry that is more complex than a shifted dipole.

\begin{figure}
   \resizebox{\hsize}{!}{\includegraphics{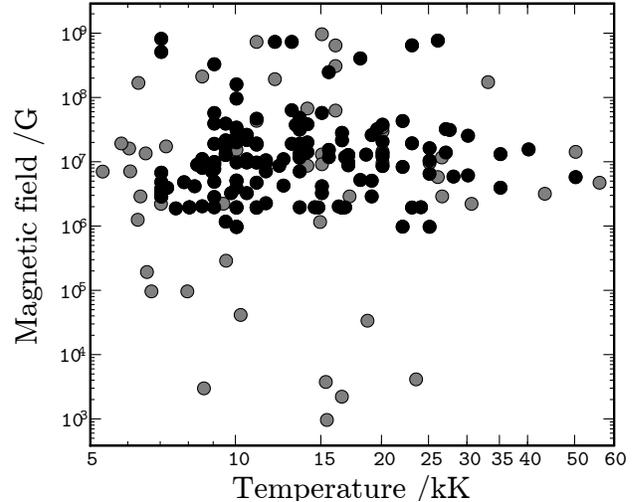}}
  \caption{Scatter plot of dipole magnetic field value vs temperature. While black dots represent all known SDSS DAHs, grey dots indicate the DAHs from literature that were not analyzed in this work. The random distribution of field strengths with respect to age indicator temperature is consistent with long decay timescale of DAHs with respect to their cooling age.}
  \label{fig:bvst}
\end{figure}

The dipole magnetic field Ohmic decay timescale is $10^{10}$ yr. Even the higher multipoles can live for such a long period of time \citep{Muslimovetal95}. Therefore, no significant correlation between temperature and magnetic field strength is expected if temperature is assumed as an indicator of age (Fig.\,\ref{fig:bvst}). This lack of correlation supports the fossil ancestry of these fields inherited from earlier stages of stellar evolution.

\section{Conclusion}
\label{Conclusion}
In this paper we have analysed 141 DAHs, 97 of them previously analysed and 44 of them being new. \citet{Gaensickeetal02} conservatively estimated that the total number of MWDs would be tripled by the complete SDSS coverage. This expectation is already surpassed before the end of the systematic search over the latest data releases. Additionally our consistent modeling over the data releases, show that within the SDSS DAH population there is a tendency to deviate from simple centered dipoles. There are clear indications of deviation from centered dipole models in least 50\% of the SDSS DAHs (see Fig\,\ref{fig:relative_std}).

\begin{figure}
   \resizebox{\hsize}{!}{\includegraphics{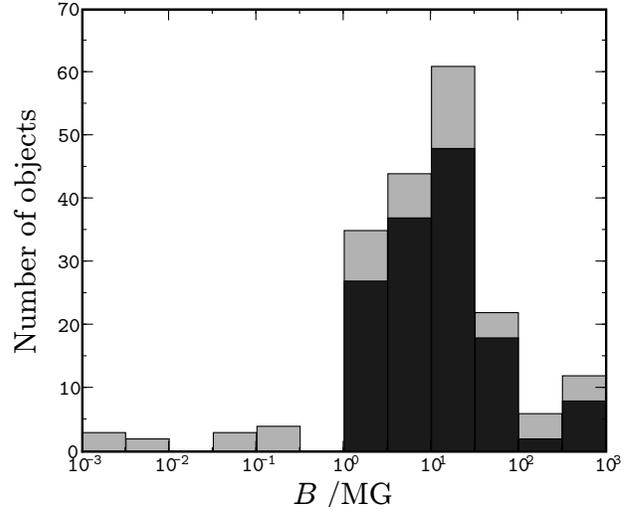}}
  \caption{Histogram of the known magnetic white dwarfs in equal intervals of $\log B$. Gray columns represent the number of all known DAHs. Black shades represent the the contribution of SDSS to DAHs.}
  \label{fig:histogram}
\end{figure}

The distribution of the magnetic field strengths of the MWDs from the \citet{Schmidtetal03} sample is concentrated in the $\sim$5\,--\,30 MG interval. We have updated the magnetic field strengths of all known DAHs and created a histogram (Fig.\,\ref{fig:histogram}. The values were taken from \citet{Jordan09}, which is an extended and corrected version of \citet{Kawkaetal07}. The same overabundance in the range $\sim$5\,--\,30 MG  as discussed in  \citet{Schmidtetal03} is apparent in Fig.\,\ref{fig:histogram}, but overall SDSS has nearly tripled the number of DAHs and conversely the completeness of the total MWD population is affected significantly by the SDSS biases because of this high impact of SDSS. 

High field MWDs are thought to be remnants of magnetic Ap and Bp stars. If flux conservation is assumed, the distribution of the polar field strengths of high field MWDs should be largest in the interval \mbox{50--500\,MG}. In our sample, objects with magnetic field strengths lower than 50 MG are more numerous than the objects with higher magnetic field strengths (see Fig.\,\ref{fig:histogram}). Part of this effect is due to our biases (see Section \ref{SDSS data}).  Nevertheless it is consistent with previous results and supports the hypothesis that magnetic fossil fields from Ap/Bp stars only are not sufficient to produce high field MWDs \citep{WickramasingheFerrario05MN}. \citet{Auriereetal07} argued that dipole magnetic field strengths of magnetic Ap/Bp stars have a ``magnetic threshold'' due to large scale stability conditions, and this results in a steep decrease in the number of magnetic Ap/Bp stars below polar magnetic fields of 300\,G. 

A possible progenitor population for MWDs with dipolar field strengths below 50 MG is the currently unobserved population of A and B stars with magnetic field strengths of 10\,--\,100\,G. \citet{WickramasingheFerrario05MN} suggested that if $\sim40\%$ of A/B stars have magnetism, this would be sufficient to explain the observed distribution of MWDs. However, the existence of this population seems to be highly unlikely since investigations of \citet{Shorlinetal02} and \citet{Bagnuloetal06} for magnetism in this population yielded null results, for median errors of 15\,--\,50\,G and 80\,G, respectively. Another candidate group for these MWDs with lower field strengths is the yet undetected magnetic F stars \citep{Schmidtetal03}. But this conclusion is strongly affected by SDSS MWD discovery biases. 

In our work we quantified the deviation from centered dipoles in our sample. To test the fossil field hypothesis one can look at the statistical properties of the fields of Ap/Bp stars. One such statistical analysis was made by \citet{Bagnuloetal02}. In their work they used the the mean longitudinal field, the crossover, the mean quadratic field and the mean field modulus to invert the magnetic field structure modeled by a dipole plus quadrupole geometry \cite[the modeling procedure is explained in depth in][]{Bagnuloetal96,Landolfietal98}. The aim of this analysis was to characterize the sample rather than find the best fit for each object. Using the aforementioned observables they analyzed 31 objects and ended up with 147 ``good fit" models. These models corresponded to the minima of the $\chi^2$ hypersurfaces in their inversion procedure. Later model parameters were weighted by these reduced $\chi^2$ values in assessing their statistical properties.

\citet{Bagnuloetal02} investigated the quadrupole vs dipole dominance of the magnetic field models by plotting weighted histograms of $B_{\rm d}/B_{\rm q}$. Where $B_{\rm d}$ is amplitude of dipole field strength ($B_p$ in this work), and $B_{\rm q}$ is the amplitude of the quadrupole field strength. The main differences between our analysis and \citet{Bagnuloetal02} was the usage of visual magnetic field distributions rather than the global magnetic field inferred from the time resolved observations. The relationship between $B_{\rm d}/B_{\rm q}$ to our skew parameter is not so straightforward since the structure of the total field depends on the angle between dipole and quadrupole components. If we were to consider $B_{\rm d}/B_{\rm q}=1$ as the point where quadrupole component starts to dominate, $63\%$ of the models can be considered dominantly quadrupolar \cite[see Fig.\,3 of][]{Bagnuloetal02}. Hence our conservative assessment of at least $50\%$ of DAHs having non dipolar field, seems to be consistent with this result.

However one needs to be careful in considering the correspondence between the geometry between MWDs and their progenitors since theoretical models expect the field on the surface to evolve under certain conditions. \citet{BraithwaiteSpruit04} investigated the stable configurations of magnetic fields in stars with their magnetohydrodynamics code. Their work resolves that initial random fields decay and within a few Alfv\'en timescales to a poloidal plus toroidal stable configuration. During the star's evolution, its toroidal field may diffuse outwards since the Ohmic diffusion time scale is within the order of the lifespan of an Ap/Bp star \citep{BraithwaiteNordlund06}. On the Ohmic time scale, the expected initial offset-dipole configuration of the surface magnetic field evolves to a simple centered dipole. Hence \citet{BraithwaiteNordlund06} hypothesized that Ap stars with centered dipole fields are likely to be older than Ap stars with non-dipolar geometries. It is important to mention that the concentration of the field inside the star is also an important parameter which contributes to the structure of the surface magnetic field. A highly concentrated field results in a surface field structure with higher order multiples after the internal toroidal field formation. The relative importance between the Ohmic diffusion (i.e. the age) and concentration of the initial field on the surface magnetic field structure was not further investigated. Nevertheless if the Ohmic diffusion timescale is effective in the lifetime of an Ap star, then the field configurations of an older population of Ap stars are more relevant for comparison to the magnetic field distribution of MWDs. This kind of work has not been undertaken for the Ap stars yet.

If we neglect the possibility that older Ap stars may behave differently from the whole sample of this group and the field structure of MWDs do not evolve unlike what \citet{Muslimovetal95} suggested (see Sec.\,\ref{General discussion}, then the global analysis of \citet{Bagnuloetal02} implicates that the distribution of field structures of chemically peculiar stars and DAHs are comparable, hence supporting the fossil field hypothesis.

In addition to the isolated evolution scenario, recently a binary star origin was proposed by \citet{Toutetal08} for generating magnetic fields in WDs. In this picture during the evolution to the cataclysmic variables (CVs) the cores of giants go to a common envelope (CE) phase. During this phase the orbital angular momentum is transferred to the envelope as the two cores spiral in together. This process causes differential rotation and convection within the CE, which are the ingredients for magnetic field generation \cite[see][and references therein]{Toutetal08}. 

A consistent account for the origins of magnetic fields in WDs awaits the finalization of the complete sample of SDSS MWDs.

\longtabL{2}{
\begin{landscape}
\begin{longtable}{lr|cc|ccc|l}
\caption{\label{fits}Model fits with centered dipole and offset dipole models with comparison to literature values. The columns indicate the SDSS name of the object; the plate, Modified Julian Date and fiber ids of the observations; the dipole magnetic field strength of the centered dipole, the inclination with respect to the line of sight of the centered dipole, the dipole magnetic field strength of the offset dipole, the inclination with respect to the line of sight of the offset dipole, the offset along the axis of the magnetic field in terms of the stellar radius, and finally comments indicate the model parameters from the literature \cite[i.e. ][]{Gaensickeetal02, Schmidtetal03, Vanlandinghametal05}}\\
\hline\hline
MWD (SDSS+) & Plate-MJD-FiberID & $B_p$ / MG  &  $i$ / deg& $B_{\rm off}$ / MG  & $z_{\rm off}$ / $r_{\rm WD}$ & $i$ / deg & fits\\
\hline
\endfirsthead
\caption{continued.}\\
\hline\hline
MWD (SDSS+) & Plate-MJD-FiberID & $B_p$ / MG  &  $i$ / deg& $B_{\rm off}$ / MG  & $z_{\rm off}$ / $r_{\rm WD}$ & $i$ / deg & Comments\\
\hline
\endhead
\hline
\endfoot
J002129.00+150223.7 & 753-52233-432 & $530.69 \pm $63.56 & $ 53.19 \pm 25.64 $ & $527.33 \pm $97.98 & $0.16 \pm $0.08 & $ 28.05 \pm 51.8 $ & 550 MG, \dots\\
J004248.19+001955.3 & 690-52261-594 & $2.00\pm0.00\footnotemark[2]$ & $ 88.22 \pm 43.37 $ & $2.58$\footnotemark[1] & $0.35 \pm $0.82 & $ 0.00\pm 23.66 $ & 14 MG, 30${}^{\circ}$\\
J021116.34+003128.5 & 405-51816-382 & $341.31 \pm $54.34 & $ 37.73 \pm 27.71 $ & $281.34 \pm $186.95 & $0.32 \pm $0.14 & $ 9.59 \pm 62.99 $ & 490 MG, \dots\\
J021148.22+211548.2 & 246-53327-048 & $166.16 \pm 7.41$ & $ 48.89 \pm 5.67 $ & $180.47 \pm 13.15$ & $0.27 \pm 0.02$ & $ 33.23 \pm 10.05 $ & 210 MG, 90${}^{\circ}$\\
J023420.63+264801.7 & 2399-53764-559 & $32.82 \pm 6.26$ & $42.08 \pm 8.09$ & $21.17 \pm 2.02$ & $0.38 \pm 0.04$ & $17.17 \pm 10.92$ & \\
J030407.40-002541.7 & 411-51817-172 & $10.95 \pm $0.98 & $ 48.57 \pm 5.72 $ & $11.13 \pm $0.97 & $0.27 \pm $0.10 & $ 51.51 \pm 17.26 $ & 11 MG, 60${}^{\circ}$\\			
J031824.19+422651.0 & 2417-53766-568 & $10.12 \pm $0.10	&$54.6\pm4.7$ & $10.77 \pm $0.10 & $0.29 \pm $0.05	&$61.1\pm10.0$ &  \\
J032628.17+052136.3 & 2339-53729-515 & $16.87 \pm 2.41$ & $53.06 \pm 26.65$ & $17.49 \pm 8.32$ & $0.34 \pm 0.12$ & $44.62 \pm 42.63$ & \\
J033145.69+004517.0 & 415-51879-378 & $13.13 \pm $1.00 & $ 49.41 \pm 38.17 $ & $12.12 \pm $9.98 & $0.21 \pm $0.02 & $ 47.43 \pm 38.49 $ & 12 MG, 60${}^{\circ}$\\			
J033320.36+000720.6 & 0415-51879-485 & $849.30 \pm 51.75$ & $50.92 \pm 8.75$ & $784.22 \pm 83.98$ & $0.19 \pm 0.05$ & $6.29 \pm 6.42$ & \\
J034308.18-064127.3 & 462-51909-117 & $9.96 \pm $2.06 & $ 41.96 \pm 9.80 $ & $9.18 \pm $2.11 & $0.23 \pm $0.04 & $ 12.59 \pm 8.55 $ & 13 MG, 45\\
J034511.11+003444.3 & 416-51811-590 & $1.96 \pm $0.42 & $ 49.01 \pm 11.56 $ & $1.46 \pm $0.36 & $0.34 \pm $0.08 & $ 16.38 \pm 6.43 $ & 1.5 MG, 0${}^{\circ}$\\			
J074850.48+301944.8 & 889-52663-507 & $6.75 \pm $0.41 & $ 40.65 \pm 7.59 $ & $8.03 \pm $0.66 & $0.40\pm $0.14 & $ 43.85 \pm 10.55 $ & 10 MG, 60${}^{\circ}$\\			
J074924.91+171355.4 & 2729-54419-282 & $13.99 \pm 1.30$ & $46.00 \pm 15.07$ & $13.62 \pm 1.86$ & $0.27 \pm 0.05$ & $44.02 \pm 53.16$ & \\
J075234.96+172525.0 & 1920-53314-106 & $10.30 \pm $1.23	&$72.4\pm22.4$ & $11.73 \pm $1.05 & $0.26 \pm $0.05	&$58.7\pm38.1$ &  \\
J075819.57+354443.7 & 757-52238-144 & $26.40\pm $3.94 & $ 37.62 \pm 22.98 $ & $32.42 \pm $5.65 & $0.39 \pm $0.07 & $ 54.51 \pm 19.44 $ & 27 MG, 30${}^{\circ}$\\
J080359.93+122944.0 & 2265-53674-033 &$40.7 \pm 2.13$& $42.97\pm8.44$ &$ 26.6\pm 11.8$&$0.26 \pm $0.04  &$17.71\pm7.21$ &\\
J080440.35+182731.0 & 2081-53357-442 & $48.47 \pm $2.93 & $ 35.67 \pm 7.59 $ & $29.26 \pm $3.18 & $0.16 \pm $0.02 & $ 33.97 \pm 8.47 $ & 49 MG, 30${}^{\circ}$\\			
J080502.29+215320.5 & 1584-52943-132 & $6.11 \pm $1.29 & $ 54.76 \pm 14.52 $ & $3.13 \pm $0.59 & $0.39 \pm $0.06 & $ 86.79 \pm 59.37 $ & 5 MG, 60${}^{\circ}$\\			
J080743.33+393829.2 & 545-52202-009 & $65.75 \pm $18.52 & $ 78.07 \pm 35.46 $ & $64.74 \pm $12.14 & $-0.01 \pm $0.01 & $ 8.25 \pm 17.26 $ & 49 MG, 30${}^{\circ}$\\		
J080938.10+373053.8 & 758-52253-044 & $39.74 \pm $5.41 & $ 41.89 \pm 14.11 $ & $30.60\pm $5.66 & $0.20\pm $0.10 & $ 15.15 \pm 75.23 $ & 40 MG, 30${}^{\circ}$\\			
J081648.71+041223.5 & 1184-52641-329 & $10.13 \pm $8.03 & $ 48.82 \pm 29.96 $ & $6.50\pm $2.73 & $0.31 \pm $0.12 & $ 0.28 \pm 27.06 $ & 10* MG, 30*${}^{\circ}$\\
J081716.39+200834.8 & 2082-53358-444 &$3.37 \pm 0.44$& $49.02\pm10.78$ &$ 3.37\pm 1.09$& $0.39 \pm $0.10 & $34.05\pm30.78$&\\
J082835.82+293448.7 & 1207-52672-635 & $33.40\pm $10.53 & $ 68.57 \pm 33.47 $ & $34.60\pm $6.91 & $0.17 \pm $0.05 & $ 50.3 \pm 36.21 $ & 30 MG, 90*${}^{\circ}$\\			
J083448.63+821059.1 & 2549 54523 135 &$14.44\pm4.57$& $54.30\pm66.45$ &$14.36\pm4.08$&$0.28\pm0.09$& $48.58\pm70.00$&\\
J083945.56+200015.7 & 2277-53705-484 & $3.38 \pm $0.49	&$48.6\pm7.7$ & $2.15 \pm $0.10 & $0.29 \pm $0.08	&$49.9\pm90\footnotemark[1]$ &  \\
J084155.74+022350.6 & 564-52224-248 & $5.00\pm $0.99 & $ 49.35 \pm 17.53 $ & $2.94 \pm $0.72 & $0.16 \pm $0.04 & $ 15.89 \pm 10.57 $ & 6 MG, 90${}^{\circ}$\\
J085106.12+120157.8 & 2430-53815-229 & $2.03 \pm $0.10	&$81.9\pm90\footnotemark[1]$ & $2.47 \pm $0.10 & $0.35 \pm $0.06	&$72.8\pm18.8$ & \\
J085523.87+164059.0 & 2431-53818-522 & $12.23 \pm $2.92	&$48.6\pm8.6$ & $7.86 \pm $1.63 & $0.36 \pm $0.06	&$10.8\pm6.1$ &  \\
J085550.67+824905.3 & 2549 54523 066 &$10.82\pm2.99 $& $54.81\pm19.36$ &$12.13\pm4.30$&$62.89\pm36.38$&$ 0.30\pm0.14 $ &\\
J085830.85+412635.1 & 830-52293-070 & $3.38 \pm $0.19 & $ 48.85 \pm 5.86 $ & $2.15 \pm $0.34 & $0.24 \pm $0.07 & $ 36.25 \pm 7.31 $ & 2 MG, 30${}^{\circ}$\\			
J090632.66+080716.0 & 1300-52973-148 & $5.98\pm3.02$ & $88.00\pm88.06$ & $5.97\pm3.13$ & $0.18\pm0.13$ & $65.13\pm82.65$ & 10 MG, 90${}^{\circ}$\\
J090746.84+353821.5 & 1212-52703-187 & $22.40\pm $8.80 & $ 48.57 \pm 16.68 $ & $11.91 \pm $3.22 & $0.26 \pm $0.08 & $ 1.77 \pm 2.72 $ & 15 MG, 60${}^{\circ}$\\			
J091005.44+081512.2 & 1300-52973-639 &$1.01\pm0.00\footnotemark[2]$& $74.46\pm42.54$ &$1.27\pm0.70$&$71.06\pm32.03$&$0.35\pm0.29$ &\\
J091124.68+420255.9 & 1200-52668-538 & $35.20\pm $5.83 & $ 35.1 \pm 39.48 $ & $18.85 \pm $4.86 & $0.23 \pm $0.06 & $ 14.23 \pm 10.72 $ & 45 MG, 60${}^{\circ}$\\			
J091437.40+054453.3 & 1193-52652-481 & $9.16 \pm $0.77 & $ 48.52 \pm 4.56 $ & $8.93 \pm $0.93 & $0.38 \pm $0.04 & $ 48.35 \pm 19.7 $ & 9.5 MG, 90${}^{\circ}$\\			
J091833.32+205536.9 & 2288-53699-547 & $2.04 \pm $0.10	&$87.2\pm41.9$ & $2.66 \pm $1.71 & $0.39 \pm $0.17	&$70.3\pm61.9$ &  \\
J092527.47+011328.7 & 475-51965-315 & $2.04\pm0.00\footnotemark[2]$ & $ 54.83 \pm 14.14 $ & $3.14 \pm $1.10 & $0.33 \pm $0.13 & $ 47.34 \pm 14.08 $ & 2.2 MG, \dots\\			
J093356.40+102215.7 & 1303-53050-525 & $2.11 \pm $0.49 & $ 72.05 \pm 22.68 $ & $2.47\pm0.00\footnotemark[2]$ & $0.37 \pm $0.12 & $ 52.39 \pm 29.38 $ & 1.5 MG, 60*${}^{\circ}$\\			
J093409.90+392759.3 & 1215-52725-241  &$1.01\pm0.00\footnotemark[2]$& $81.18\pm23.30$ &$ 1.35\pm 0.25$& $57.43\pm29.65$ &$ 0.39\pm $0.13 & \\
J093447.90+503312.2 & 901-52641-373 & $7.35 \pm $2.21 & $ 54.61 \pm 37.93 $ & $4.29 \pm $1.06 & $0.28 \pm $0.07 & $ 46.87 \pm 49.83 $ & 9.5 MG, 60${}^{\circ}$\\
J094235.02+205208.3 & 2292-53713-019 & $39.21 \pm 4.55$ & $3.73 \pm 1.64$ & $37.94 \pm 9.07$ & $0.04 \pm 0.02$ & $1.24 \pm 13.74$ & \\
J094458.92+453901.2 & 1202-52672-577 & $15.91 \pm $9.10 & $ 68.35 \pm 44.53 $ & $17.41 \pm $7.12 & $0.34 \pm $0.14 & $ 61.00 \pm 90\footnotemark[1] $ & 14 MG, 90${}^{\circ}$\\			
J100005.67+015859.2 & 500-51994-557 & $19.74 \pm $10.26 & $ 48.5 \pm 33.13 $ & $8.10\pm $3.23 & $0.34 \pm $0.13 & $ 4.21 \pm 90\footnotemark[1] $ & 20 MG, 30${}^{\circ}$\\			
J100356.32+053825.6 & 996-52641-295 & $672.07 \pm $118.63 & $ 40.8 \pm 34.86 $ & $667.67 \pm $131.71 & $0.08 \pm $0.05 & $ 36.67 \pm 23.11 $ & 900* MG, \dots\\
J100657.51+303338.1 & 1953-53358-415 & $1.00 \pm $0.10	&$82.5\pm30.8$ & $1.30 \pm $1.23 & $-0.37 \pm $0.39	&$14.3\pm13.1$ &  \\
J100715.55+123709.5 & 1745-53061-313 & $5.41 \pm $67.28 & $ 68.6 \pm 31.88 $ & $6.01 \pm $2.55 & $0.40\pm $0.22 & $ 50.7 \pm 90\footnotemark[1] $ & 7 MG, 60${}^{\circ}$\\			
J100759.80+162349.6 & 2585 54097 030 &$19.18\pm3.36$&$42.02\pm17.48$&$13.50\pm2.92$&$-0.22\pm0.05$ & $33.31\pm21.38$&\\
J101428.09+365724.3 & 1954-53357-393 & $11.09 \pm 1.50$ & $48.75 \pm 11.72$ & $12.14 \pm 2.71$ & $0.05 \pm 0.04$ & $56.81 \pm 34.40$ & \\
J101529.62+090703.8 & 1237-52762-533 & $4.09 \pm $0.86 & $ 49 \pm 15.27 $ & $1.98 \pm $0.35 & $0.26 \pm $0.08 & $ 1.48 \pm 60.69 $ & 5* MG, 90*${}^{\circ}$\\			
J101618.37+040920.6 & 574-52355-166 & $2.01\pm0.00\footnotemark[2]$ & $ 88.16 \pm 104.01 $ & $7.24 \pm $4.77 & $0.19 \pm $0.16 & $ 60.55 \pm 63.4 $ & 7.5 MG, 30${}^{\circ}$\\			
J101805.04+011123.5 & 503-51999-244 & $99.92\pm5.90$ & $27.24\pm6.16 $ & $39.87 \pm $2.25 & $0.26 \pm 0.02$ & $14.85 \pm 3.17 $ & 120 MG, \dots\\
J102220.69+272539.8 & 2350-53765-543 &$4.91\pm0.31 $& $27.58\pm30.65$ &$ 5.79\pm2.82 $& $ 0.39\pm0.27 $ & $54.97\pm35.63$&\\
J102239.06+194904.3 & 2374-53765-544 & $2.94 \pm $0.71	&$49.0\pm13.0$ & $3.87 \pm $1.11 & $0.40 \pm $0.07	&$51.3\pm40.3$ & \\
J103532.53+212603.5 & 2376-53770-534 &$2.96\pm 0.33$& $49.01\pm9.51$ &$ 1.515\pm 0.313$&  $ 0.39\pm0.05 $& $0.00\pm57.44$&\\
J105404.38+593333.3 & 561-52295-008 & $17.41 \pm $7.90 & $ 90.0 \pm 25.54 $ & $17.60 \pm $10.22 &  $ 0.16 \pm 0.05 $& $67.59 \pm $59.48 & 17 MG, 90${}^{\circ}$\\			
J105628.49+652313.5 & 490-51929-205 & $29.27 \pm $5.78 & $ 41.74 \pm 15.24 $ & $20.80\pm $4.10 & $0.29 \pm $0.54 & $ 5.27 \pm 12.8 $ & 28 MG, 60${}^{\circ}$\\			
J105709.81+041130.3 & 580-52368-274 &$2.03\pm0.00\footnotemark[2]$& $48.99\pm7.70$  &$2.48\pm0.0$& $41.40\pm13.13$  &$0.38\pm0.06$ &  \\
J111010.50+600141.4 & 950-52378-568 & $6.37 \pm $2.32 & $ 68.6 \pm 16.26 $ & $6.71 \pm $1.58 & $0.39 \pm $0.08 & $ 49.08 \pm 18.99 $ & 6.5 MG, 70${}^{\circ}$\\			
J111812.67+095241.3 & 1222-52763-477 & $3.38 \pm $0.72 & $ 48.14 \pm 32.44 $ & $2.67 \pm $0.60 & $0.40\pm $0.10 & $ 51.09 \pm 43.5 $ & 6 MG, 60*${}^{\circ}$\\		
J112030.34-115051.1 & 2874-54561-512 & $8.90 \pm 1.02$ & $50.37 \pm 20.24$ & $7.72 \pm 1.24$ & $0.35 \pm 0.10$ & $38.69 \pm 34.08$ & \\
J112257.10+322327.8 & 1979-53431-512 & $11.38 \pm $3.42	&$49.0\pm12.3$ & $7.46 \pm $1.68 & $0.37 \pm $0.11	&$4.2\pm6.3$ &  \\
J112328.49+095619.3 & 1222-52763-625 &$1.21\pm\footnotemark[2]$& $81.18\pm17.28$ &$ 1.50\pm 0.36$& $0.39\pm 0.06$ &$ 50.93\pm90\footnotemark[1]$ &\\
J112852.88-010540.8 & 326-52375-565 & $2.00\pm0.00\footnotemark[2]$ & $ 89.4 \pm 85.15 $ & $2.30\pm $5.02 & $0.20\pm $0.14 & $ 23.15 \pm 30.18 $ & 3 MG, 60${}^{\circ}$\\		
J112926.23+493931.8 & 966-52642-474 & $5.31 \pm $0.64 & $ 48.72 \pm 6.34 $ & $2.43 \pm $0.33 & $0.39 \pm $0.09 & $ 7.04 \pm 12.45 $ & 5 MG, 60${}^{\circ}$\\
J113215.38+280934.3 & 2219-53816-329 &$3.01\pm0.82$&$49.00\pm17.43$ &$2.84\pm0.43$&$0.39\pm0.049$ &$18.60\pm5.56$&\\
J113357.66+515204.8 & 879-52365-586 & $8.64 \pm $0.78 & $ 74.55 \pm 19.87 $ & $7.69 \pm $0.69 & $0.39 \pm $0.04 & $ 47.91 \pm 19.99 $ & 7.5 MG, 90${}^{\circ}$\\		
J113756.50+574022.4 & 1311-52765-421 & $5.00\pm $0.34 & $ 33.45 \pm 9.92 $ & $2.83 \pm $0.25 & $0.17 \pm $0.01 & $ 0.07 \pm 0.06 $ & 9 MG, 60*${}^{\circ}$\\		
J113839.51-014903.0 & 327-52294-583 & $22.71 \pm $1.26 & $ 54.31 \pm 14.41 $ & $24.09 \pm $1.92 & $0.21 \pm $0.04 & $ 50.16 \pm 61.4 $ & 24 MG, 60${}^{\circ}$\\		
J114006.37+611008.2 & 776-52319-042 & $50.19 \pm $17.78 & $ 21.8 \pm 42.46 $ & $52.82 \pm $8.82 & $0.04 \pm $0.02 & $ 36.88 \pm 50.59 $ & 58 MG, 20${}^{\circ}$\\		
J114829.00+482731.2 & 1446-53080-324 & $32.47 \pm $7.11 & $ 80.89 \pm 45.93 $ & $32.44 \pm $2.70 & $0.17 \pm $0.09 & $ 69.98 \pm 33.06 $ & 33 MG, 90${}^{\circ}$\\		
J115418.14+011711.4 & 515-52051-126 & $33.47 \pm $2.07 & $ 88.14 \pm 23.62 $ & $25.72 \pm $2.51 & $0.10\pm $0.03 & $ 76.03 \pm 32.13 $ & 32 MG, 45	\\		
J115917.39+613914.3 & 777-52320-69 & $20.10\pm $6.70 & $ 50.63 \pm 62.9 $ & $10.14 \pm $6.89 & $0.30\pm $0.07 & $ 50.63 \pm 62.9 $ & 15.5 MG, 60${}^{\circ}$\\	
J120150.10+614257.0 & 778-52337-264 & $11.35 \pm $1.53 & $ 28.04 \pm 10.98 $ & $7.60\pm $0.87 & $0.16 \pm $0.02 & $ 18.59 \pm 12.62 $ & 20 MG, 90${}^{\circ}$\\	
J120609.80+081323.7 & 1623-53089-573 & $760.63 \pm $281.66 & $ 30.74 \pm 36.38 $ & $312.24 \pm $73.90 & $0.24 \pm $0.13 & $ 25.69 \pm 24.66 $ & 830* MG, \dots	\\		
J120728.96+440731.6 & 1369-53089-048 & $2.03\pm0.00\footnotemark[2]$ & $ 73.01 \pm 29.93 $ & $2.07\pm0.00\footnotemark[2]$ & $-0.27 \pm $0.59 & $ 10.16 \pm 21.53 $ & 2.5 MG, 90${}^{\circ}$\\	
J121209.31+013627.7 & 518-52282-285 & $10.12 \pm $0.93 & $ 48.8 \pm 4.69 $ & $10.38 \pm $1.02 & $0.29 \pm $0.05 & $ 40.65 \pm 13.27 $ & 13 MG, 80${}^{\circ}$\\		
J121635.37-002656.2 & 288-52000-276 & $59.70\pm $10.23 & $ 37.85 \pm 37.52 $ & $44.57 \pm $7.18 & $-0.06 \pm $0.01 & $ 20.15 \pm 30.89 $ & 61 MG, 90${}^{\circ}$\\		
J122209.44+001534.0 & 289-51990-349 & $14.70\pm $4.70 & $ 82.52 \pm 90\footnotemark[1] $ & $16.27 \pm $11.39 & $0.24 \pm $0.31 & $ 82.52 \pm 90\footnotemark[1] $ & 14 MG, 80${}^{\circ}$\\		
J122249.14+481133.1 & 1451-53117-582 & $8.05 \pm $2.24 & $ 56.67 \pm 13.05 $ & $8.70\pm $3.90 & $-0.38 \pm $0.08 & $ 53.05 \pm 18.28 $ & 8 MG, 90*${}^{\circ}$\\	
J122401.48+415551.9 & 1452-53112-181 & $22.36 \pm $3.02 & $ 66.08 \pm 23.71 $ & $24.75 \pm $5.76 & $0.26 \pm $0.06 & $ 58.28 \pm 23.89 $ & 23* MG, 60${}^{\circ}$\\	
J123414.11+124829.6 & 1616-53169-423 & $4.32 \pm $0.27 & $ 49.35 \pm 9.82 $ & $2.40\pm $0.56 & $0.40\pm $0.05 & $ 9.99 \pm 2.72 $ & 7 MG, 60*${}^{\circ}$\\	
J124806.38+410427.2 & 1456-53115-190 & $7.03 \pm $1.19 & $ 48.81 \pm 9.27 $ & $3.61 \pm $0.47 & $0.29 \pm $0.06 & $ 5.5 \pm 10.08 $ & 8 MG, 90${}^{\circ}$\\	
J124836.31+294231.2 & 2457-54180-112 & $3.95 \pm 0.25$ & $48.93 \pm 4.93$ & - & - &  - & \\
J124851.31-022924.7 & 337-51997-264 & $7.36 \pm $2.19 & $ 48.9 \pm 10.11 $ & $8.01 \pm $1.46 & $0.40\pm $0.09 & $ 49.97 \pm 27.39 $ & 7 MG, 40${}^{\circ}$\\	
J125044.42+154957.4 & 1770-53171-530 & $20.71 \pm $3.66 & $ 56.7 \pm 11.21 $ & $20.30\pm $2.43 & $0.12 \pm $0.02 & $ 53.83 \pm 38.18 $ & 20 MG, 60${}^{\circ}$\\	
J125416.01+561204.7 & 1318-52781-299 & $38.86 \pm $9.03 & $ 20.97 \pm 40.41 $ & $36.98 \pm $9.20 & $0.01 \pm 1e+99$  & $ 10.25 \pm 21.24 $ & 52 MG, 30${}^{\circ}$\\	
J125434.65+371000.1 & 1989-53772-041 & $4.10 \pm $0.35	&$41.9\pm19.2$ & $4.89 \pm $0.42 & $0.40 \pm $0.03	&$50.1\pm21.0$ & \\
J125715.54+341439.3 & 2006-53476-332 & $11.45 \pm $0.71	&$0.5\pm0.6$ & $13.70 \pm $1.69 & $0.07 \pm $0.02	&$7.7\pm12.0$ &  \\
J132002.48+131901.6 & 1773-53112-011 & $2.02\pm0.00\footnotemark[2]$ & $ 88.21 \pm 80.46 $ & $2.64 \pm $4.88 & $-0.38 \pm $0.52 & $ 6.99 \pm 35.73 $ & 5 MG, 60${}^{\circ}$\\	
J133340.34+640627.4 & 603-52056-112 & $10.71 \pm $1.03 & $ 50.29 \pm 11.45 $ & $13.81 \pm $1.36 & $0.38 \pm $0.07 & $ 55.7 \pm 24.63 $ & 13 MG, 60${}^{\circ}$\\	
J134043.10+654349.2 & 497-51989-182 & $4.32 \pm $0.76 & $ 49.04 \pm 9.75 $ & $5.28 \pm $1.07 & $0.40\pm $0.06 & $ 34.03 \pm 6.8 $ & 3 MG, 60${}^{\circ}$\\	
J134820.79+381017.2 & 2014-53460-236 & $13.65 \pm $2.66	&$89.4\pm90\footnotemark[1]$ & $14.45 \pm $4.65 & $0.22 \pm $0.04	&$54.8\pm25.3$ &  \\
J135141.13+541947.4 & 1323-52797-293 & $761.00\pm $56.42 & $ 74.18 \pm 21.65 $ & $772.73 \pm $94.41 & $-0.03 \pm $0.01 & $ 68.25 \pm 62.04 $ & 760 MG, 20${}^{\circ}$\\		
J140716.66+495613.7 & 1671-53446-453 & $12.49 \pm $6.20	&$88.1\pm90\footnotemark[1]$ & $13.20 \pm $4.21 & $0.24 \pm $0.10	&$63.3\pm81.1$ & \\
J141906.19+254356.5 & 2131-53819-317 & $2.03 \pm $0.10	&$81.2\pm8.7$ & $2.56 \pm $0.10 & $0.38 \pm $0.03	&$54.8\pm10.4$ & \\
J142703.40+372110.5 & 1381-53089-182 & $27.04 \pm $3.20 & $ 56.89 \pm 12.61 $ & $30.59 \pm $2.34 & $0.20\pm $0.02 & $ 6.39 \pm 22.5 $ & 30 MG, 60*${}^{\circ}$\\	
J143019.05+281100.8 & 2134-53876-423 & $9.34 \pm $1.44	&$5.6\pm4.5$ & $6.25 \pm $0.75 & $0.16 \pm $0.03	&$5.6\pm4.5$ &  \\
J143218.26+430126.7 & 1396-53112-338 & $1.01\pm0.00\footnotemark[2]$& $ 78.3 \pm 90\footnotemark[1] $ & $1.48 $\footnotemark[1] & $-0.39\pm $0.56 & $ 55.37 \pm 81.19 $ & 1.5 MG, 90${}^{\circ}$\\	
J143235.46+454852.5 & 1288-52731-449 & $12.29 \pm $6.98 & $ 50.62 \pm 88.7 $ & $13.08 \pm $4.78 & $0.40\pm $0.11 & $ 25.84 \pm 33.41 $ & 10 MG, 30${}^{\circ}$\\
J144614.00+590216.7 & 608-52081-140 & $4.42 \pm $3.79 & $ 48.9 \pm 31.58 $ & $2.18\pm0.00\footnotemark[2]$ & $0.40\pm $0.26 & $ 4.72 \pm 13.13 $ & 7 MG, 70${}^{\circ}$\\
J145415.01+432149.5 & 1290-52734-469 & $2.35 \pm $0.88 & $ 49.04 \pm 17.87 $ & $2.15 \pm $1.09 & $0.27 \pm $0.11 & $ 53.38 \pm 22.89 $ & 5 MG, \dots	\\
J150813.20+394504.9 & 1398-53146-633 & $13.23 \pm $3.11 & $ 37.55 \pm 4.17 $ & $7.96 \pm $0.76 & $0.40\pm $0.04 & $ 24.68 \pm 5.61 $ & 20 MG, 90${}^{\circ}$\\	
J151130.17+422023.0 & 1291-52735-612 & $22.40 \pm $9.41 &$48.6\pm19.5$ & $8.37 \pm $1.07 & $0.31 \pm $0.06 &$5.8\pm21.1$ &  12 MG, 60${}^{\circ}$\\
J151415.65+074446.5 & 1817-53851-534 &$35.34\pm2.80$&$56.59\pm17.23$&$35.88\pm3.75$&$0.12\pm0.043$ & $56.08\pm22.18$ &\\
J151745.19+610543.6 & 613-52345-446 & $13.98 \pm $7.36 & $ 60.07 \pm 196.18 $ & $6.07 \pm $3.24 & $0.23 \pm $0.35 & $ 24.86 \pm 47.72 $ & 17 MG, 30${}^{\circ}$\\	
J152401.60+185659.2 & 2794-54537-410 & $11.96 \pm 1.85$ & $41.53 \pm 8.79$ & $10.27 \pm 3.12$ & $0.11 \pm 0.02$ & $33.40 \pm 6.58$ & \\
J153532.25+421305.6 & 1052-52466-252 & $5.27 \pm $4.05 & $ 0.35 \pm 41.68 $ & $9.05 \pm $5.27 &   $-0.29 \pm 0.12$& $ 0.13 \pm 0.37 $ & 4.5 MG, 60${}^{\circ}$\\		
J153829.29+530604.6 & 795-52378-637 & $13.99 \pm $3.82 & $ 48.96 \pm 15.69 $ & $15.99 \pm $3.03 & $0.36 \pm $0.11 & $ 53.36 \pm 29.05 $ & 12 MG, 30${}^{\circ}$\\	
J153843.10+084238.2 & 1725 54266 297 &$13.20\pm4.34$& $41.09\pm90\footnotemark[1]$ &$9.63\pm$&$24.98\pm34.53$&$ 0.33\pm0.08 $ &\\
J154213.48+034800.4 & 594-52045-400 & $8.35 \pm $2.60 & $ 54.34 \pm 35.5 $ & $8.25 \pm $2.72 & $0.19 \pm $0.16 & $ 44.48 \pm 30.63 $ & 8 MG, 60${}^{\circ}$\\	
J154305.67+343223.6 & 1402-52872-145 & $4.09 \pm 2.67$ & $62.06 \pm 158.24$ & $4.02 \pm 1.09$ & $0.20 \pm 0.06$ & $62.67 \pm 22.69$ & \\
J160437.36+490809.2 & 622-52054-330 & $59.51 \pm $4.64 & $ 40.83 \pm 28.87 $ & $38.36 \pm $3.61 & $0.14 \pm $0.02 & $ 7.95 \pm 10.24 $ & 53 MG, 0${}^{\circ}$\\	
J164357.02+240201.3 & 1414-53135-191 & $2.00\pm0.00\footnotemark[2]$ & $ 88.01 \pm 56.93 $ & $2.41 \pm $10.10 & $-0.34 \pm $0.99 & $ 18.54 \pm 33.29 $ & 4 MG, 90${}^{\circ}$\\	
J164703.24+370910.3 & 818-52395-026 & $2.10\pm $0.67 & $ 68.39 \pm 18.11$ & $2.15 \pm $1.21 & $0.28 \pm $0.10 & $ 72.53 \pm 18.41$ & 2* MG, 90*${}^{\circ}$\\									
J165029.91+341125.5 & 1175-52791-482 & $3.38 \pm $0.67 & $ 48.78 \pm 13.64 $ & $3.92 \pm $1.05 & $0.40\pm $0.12 & $ 57.74 \pm 32.5 $ & 3* MG, 	0*${}^{\circ}$\\
J165203.68+352815.8 & 820-52438-299 & $7.37 \pm $2.92 & $ 48.98 \pm 11.77 $ & $9.53 \pm $3.53 & $0.40\pm $0.08 & $ 51.15 \pm 30.36 $ & 9.5 MG, 60${}^{\circ}$\\	
J165249.09+333444.9 & 1175-52791-095 & $5.07 \pm 4.18$ & $54.25 \pm 28.04$ & $5.82 \pm 1.90$ & $0.39 \pm 0.11$ & $49.85 \pm 36.73$ & \\
J170400.01+321328.7 & 976-52413-319 & $50.11 \pm $25.08 & $ 54.87 \pm 118.04 $ & $56.16 \pm $8.84 & $0.27 \pm $0.05 & $ 0.34 \pm 53.88 $ & 5 MG, 90*${}^{\circ}$\\	
J171556.29+600643.9 & 354-51792-318 & $2.03 \pm0.00\footnotemark[2]$ & $ 86.63 \pm 31.39 $ & $2.03\pm0.00\footnotemark[2]$ & $-0.19 \pm $0.09 & $ 60.85 \pm 35.29 $ & 4.5 MG, 60${}^{\circ}$\\	
J172045.37+561214.9 & 367-51997-461 & $19.79 \pm $5.42 & $ 56.7 \pm 21.33 $ & $9.72 \pm $2.68 & $0.31 \pm $0.18 & $ 0.85 \pm 36.62 $ & 7 MG, 30${}^{\circ}$\\		
J172329.14+540755.8 & 359-51821-415 & $32.85 \pm $3.56 & $ 37.43 \pm 26.52 $ & $28.86 \pm $2.71 & $0.04 \pm $0.03 & $ 25.4 \pm 27.51 $ & 35 MG, 10${}^{\circ}$\\		
J172932.48+563204.1 & 358-51818-239 & $27.26 \pm $7.04 & $ 51.19 \pm 67.41 $ & $27.17 \pm $18.71 & $0.22 \pm $0.10 & $ 41.74 \pm 90$\footnotemark[1] & 35 MG, \dots	\\		
J202501.10+131025.6 & 2257-53612-167 & $10.10 \pm $1.76 &$68.5\pm9.1$ & $10.72 \pm $1.71 & $0.29 \pm $0.04 &$53.7\pm9.0$ &  \\
J204626.15-071037.0 & 635-52145-227 & $2.03 \pm0.00\footnotemark[2]$ & $ 49.33 \pm 25.34 $ & $2.62 \pm $0.53 & $0.40\pm $0.06 & $ 54.97 \pm 17.54 $ & 2 MG, 60${}^{\circ}$\\
J205233.52-001610.7 & 982-52466-019 & $13.42 \pm $3.73 & $ 68.57 \pm 12.86 $ & $1.31 \pm $0.23 & $0.13 $0.03 & $ 56.97 \pm 16.83 $ & 13 MG, 80${}^{\circ}$\\			
J214900.87+004842.8 & 1107-52968-374 & $10.09 \pm $4.71 & $ 46.97 \pm 99.7 $ & $5.53 \pm $2.90 & $0.22 \pm $0.11 & $ 5.00 \pm 9.13 $ & 10 MG, 60${}^{\circ}$\\		
J214930.74-072812.0 & 644-52173-350 & $44.71 \pm $1.92 & $ 67.27 \pm 26.38 $ & $44.71 \pm $2.80 & $0.06 \pm $0.05 & $ 60.11 \pm 16.52 $ & 42 MG, 30${}^{\circ}$\\	
J215148.31+125525.5 & 733-52207-522 & $20.76 \pm $1.39 & $ 68.3 \pm 12.21 $ & $21.80\pm $3.09 & $0.12 \pm $0.03 & $ 68.66 \pm 19.04 $ & 21 MG, 90${}^{\circ}$\\	
J220435.05+0012 42.9 & 372-52173-626 & $1.02 \pm $0.10	&$71.2\pm90\footnotemark[1]$ & $2.50 \pm $5.47 & $-0.36 \pm $0.69	&$3.1\pm13.6$ &  \\
J221828.59-000012.2 & 374-51791-583 & $257.54\pm48.71$ & $17.68\pm17.63$ & $212.18\pm34.78$& $0.06\pm0.02$ & $17.09\pm27.01$ & 225 MG, 30${}^{\circ}$\\
J224741.46+145638.8 & 740-52263-444 & $42.11 \pm 2.83$ & $ 53.02 \pm 9.70 $ & $46.95 \pm 4.30$ & $-0.17  \pm 0.02$ & $ 33.16 \pm 9.84 $ & 560 MG, \dots	\\
J225726.05+075541.7 & 2310-53710-420 & $16.17 \pm $2.81	&$74.9\pm16.1$ & $17.39 \pm $3.21 & $0.15 \pm $0.05	&$78.5\pm34.8$ &  \\
J231951.73+010909.3 & 382-51816-565 & $9.35 \pm $31.50 & $ 48.98 \pm 11.36 $ & $6.06 \pm $1.24 & $0.40\pm $0.09 & $ 12.48 \pm 50.66 $ & 1.5* MG, 90*${}^{\circ}$\\		
J232248.22+003900.9 & 383-51818-421 & $21.40\pm $3.36 & $ 49.00 \pm 15.42 $ & $21.65 \pm $4.48 & $0.29 \pm $0.07 & $ 44.174 \pm 25.58 $ & 19 MG, 60${}^{\circ}$\\										
J234605.44+385337.6 & 1883-53271-272 & $798.1\pm163.6$ & $2.50\pm1.07$ & $706.0 \pm 238.9$ & $0.12\pm0.06$ & $86.6\pm15.4$ &1000* MG, \dots\\
J234623.69-102357.0 & 648-52559-142 & $9.17 \pm $1.58 & $ 48.78 \pm 9.62 $ & $2.25 \pm $0.29 & $0.39 \pm $0.07 & $ 8.42 \pm 2.6 $ & 2.5 MG, 90*${}^{\circ}$ \\
\label{tab:results}
\end{longtable}
Asterisks indicate field strengths or inclinations with uncertainties greater than 10\% in \citet{Vanlandinghametal05}. \\
${}^1$ The errors are very large. \\
${}^2$ The errors are very small. 
\end{landscape}
}

\begin{acknowledgements}
      Part of this work supported by the DLR under grant 50 OR 0201.
      The Sloan Digital Sky Survey (SDSS) is a joint project of the University of Chicago, Fermilab, the Institute for Advanced Study, the Japan Participation Group, the Johns Hopkins University, the Max-Planck-Institut f\"ur Astronomie, New Mexico State University, Princeton University, the United States Naval Observatory, and the University of Washington. Apache Point Observatory, site of the SDSS, is operated by the Astrophysical Research Consortium. Funding for the project has been provided by the Alfred P. Sloan Foundation, the SDSS member institutions, the National Aeronautics and Space Administration, the National Science Foundation, the U.S. Department of Energy, Monbusho, and the Max Planck Society. The SDSS World Wide Web site is http://www.sdss.org/.

	B.K. would like to acknowledge support from the International Max-Planck Research School (IMPRS) for Astronomy and Cosmic Physics at the University of Heidelberg, and Heidelberg Graduate School of Fundamental Physics (HGSFP).
\end{acknowledgements}

\bibliographystyle{aa}


\Online
\begin{appendix} 
\section{Figures of all fits not shown in the printed version}
The full preprint version can download from \href{http://www.ari.uni-heidelberg.de/mitarbeiter/bkulebi/papers/12570_online.pdf}{http://www.ari.uni-heidelberg.de/mitarbeiter/bkulebi/papers/12570$\_$online.pdf}.
\end{appendix}
\end{document}